\documentclass[a4paper,aps,prd,10pt,preprintnumbers,twocolumn,superscriptaddress,nofootinbib,amsmath,amssymb]{revtex4-1}
\usepackage{graphicx,xcolor}
\usepackage[utf8]{inputenc}
\usepackage[T1]{fontenc}
\usepackage{cmap}

\newcommand{\cL}{\mathcal{L}}

\newcommand{\Schw}{Schwarzschild}

\newcommand{\beq}{\begin{equation}}
\newcommand{\eeq}{\end{equation}}
\newcommand{\bea}{\begin{eqnarray}}
\newcommand{\eea}{\end{eqnarray}}

\def\lin{ --- }

\def\der{|}

\def\sun{{\rm sun}}

\def\d{\dif}

\def\tt{\theta}

\def\der{|}

\providecommand{\dif}{\mathrm{d}} \def\d{\dif}

\def\mir{\mathrm{r}}
\def\mit{\mathrm{\theta}}
\def\mip{\mathrm{\phi}}

\graphicspath{{pic/}}

\par
\par

\begin{document}
\title{
String loop vibration around Schwarzschild black hole
}

\author{Mariia Churilova}\email{wwrttye@gmail.com}
\affiliation{Research Centre for Theoretical Physics and Astrophysics, Institute of Physics, Silesian University in Opava, CZ-74601 Opava, Czech Republic}
\author{Martin Kolo\v{s}}\email{martin.kolos@physics.slu.cz}
\affiliation{Research Centre for Theoretical Physics and Astrophysics, Institute of Physics, Silesian University in Opava, CZ-74601 Opava, Czech Republic}
\author{Zden\v{e}k Stuchl{\'i}k}\email{zdenek.stuchlik@physics.slu.cz}
\affiliation{Research Centre for Theoretical Physics and Astrophysics, Institute of Physics, Silesian University in Opava, CZ-74601 Opava, Czech Republic}
\begin{abstract}
%
String loop vibrations in a central plane of a Schwarzschild black hole are investigated for various string equations of state. We discuss string loop stability and derive frequencies of vibrational modes. Using the vibrating string loop model we fit the quasi-periodic oscillation (QPO) observed in X-ray signal coming from some compact sources. We demonstrate how the string-loop parameters are related to the radial and vertical fundamental vibration modes, and how the vibrational instability can be related to the Q-factor characterizing the observed QPOs.

%
%
\end{abstract}

\maketitle


\section{Motivations} \label{intro}


Long-range forces provided by gravitational and electromagnetic (EM) interactions play a vital role in comprehending the proper understanding of high-energy processes around black holes (BHs) \cite{Stu-etal:2020:Universe:}. Observational evidence strongly supports the existence of magnetic fields (MFs) in the vicinity of BHs, emphasizing their significance in these processes \cite{Eatough-etal:2013:Natur:,Daly:2019:APJ:}. Orders of magnitude of MFs around BHs may vary from a few Gs up to $10^8$~Gs and more, depending on the source generating the field. For stellar-mass BHs observed in X-ray binaries, the characteristic strength of MFs is of order $10^8$~Gs, while for supermassive BHs (SMBHs), the characteristic strength is of order $10^4$~Gs. In realistic astrophysical scenarios, the energy densities of MFs at such magnitudes are insufficient to significantly influence the geometry of the background spacetime - the spacetime metric around a BH can be accurately described by the Kerr or Schwarzschild solutions of the Einstein field equations.

In the vicinity of BH where the gravitational and electromagnetic field plays an important role, relativistic magnetohydrodynamics (GRMHD) should be used. It was shown by V.~S.~Semenov and his colleagues \cite{Sem-Ber:1990:ASS:} that there is formal equivalence between GRMHD equations and equations for dynamics of relativistic string. Originally, the cosmic strings were introduced as remnants of the phase transitions in the very early universe \cite{Vil-She:1994:CSTD:} or strings represented as superconducting vortices were considered in \cite{Wit:1985:NuclPhysB:}. In the case of the cosmic strings general relativistic effect of "conical reduction" of the spacetime occurs -- behavior of such strings in relation to black holes was studied e.g. by Larsen \cite{Larsen:1994:CQG:}. For the MHD inspired string loops considered in the present paper, the GR effects are irrelevant regarding their internal structure but influenced significantly the external conditions. However, the current-carrying string loops could represent also plasma exhibiting a string-like behavior due to dynamics of the magnetic field lines \cite{Sem-Dya-Pun:2004:Sci:}, or due to the thin flux tubes of magnetized plasma simply described as 1D strings \cite{Sem-Ber:1990:ASS:}. Within this project, we would like to explore connections between complex magnetic flux tube evolution (1D plasma dynamic) and a more simple string loop model.

\begin{figure}
\centering
\includegraphics[width=0.9\hsize]{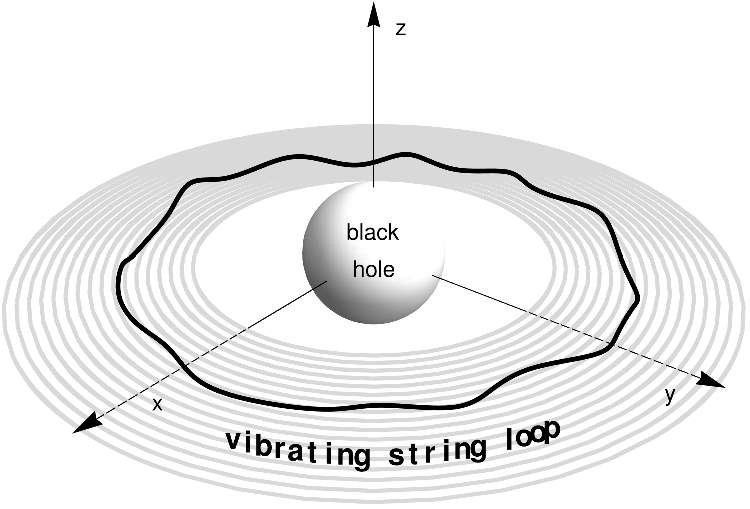}
\caption{ Schematic representation of vibrating circular string loop in equatorial plane of Schwarzschild BH. 
\label{figLoop}
}
\end{figure}

In this article, we are studying circular string structures (current-carrying with scalar field, "dust" particle loop) threaded onto the symmetry axis of the BH, see Fig.~\ref{figLoop}. The string loop can oscillate, changing its radius in the plane perpendicular to the symmetry axis, while propagating along the symmetry axis \cite{Kol-Stu:2013:PHYSR4:}. The axisymmetric string loops are governed by their tension and angular momentum. The tension of the string loops prevents their expansion beyond some radius, while their worldsheet current introduces an angular momentum preventing them from collapsing. String loop vibration traveling along the loop can also generate angular moment on the sting loop even for simple Nambu–Goto strings (without a scalar field). It has been shown that frequencies of the string loop quasi-harmonic oscillations (QPOs) can fit twin QPOs observed with resonant frequency ratio $3:2$ in three Galactic microquasars \cite{Stu-Kol:2014:PHYSR4:}. Resonances and higher frequency string loop vibrational modes and their relation to observed QPOs will be also explored. The oscillating string loops can have frequencies similar to frequencies of epicyclic geodesic motion \cite{Stu-Kot-Tor:2013:ASTRA:}, or epicyclic oscillatory motion of charged particles around magnetized black holes \cite{Kol-Stu-Tur:2015:CLAQG:}.

\section{String loop dynamics around black hole} \label{SecVEFF}

\begin{figure*}
\includegraphics[width=0.95\hsize]{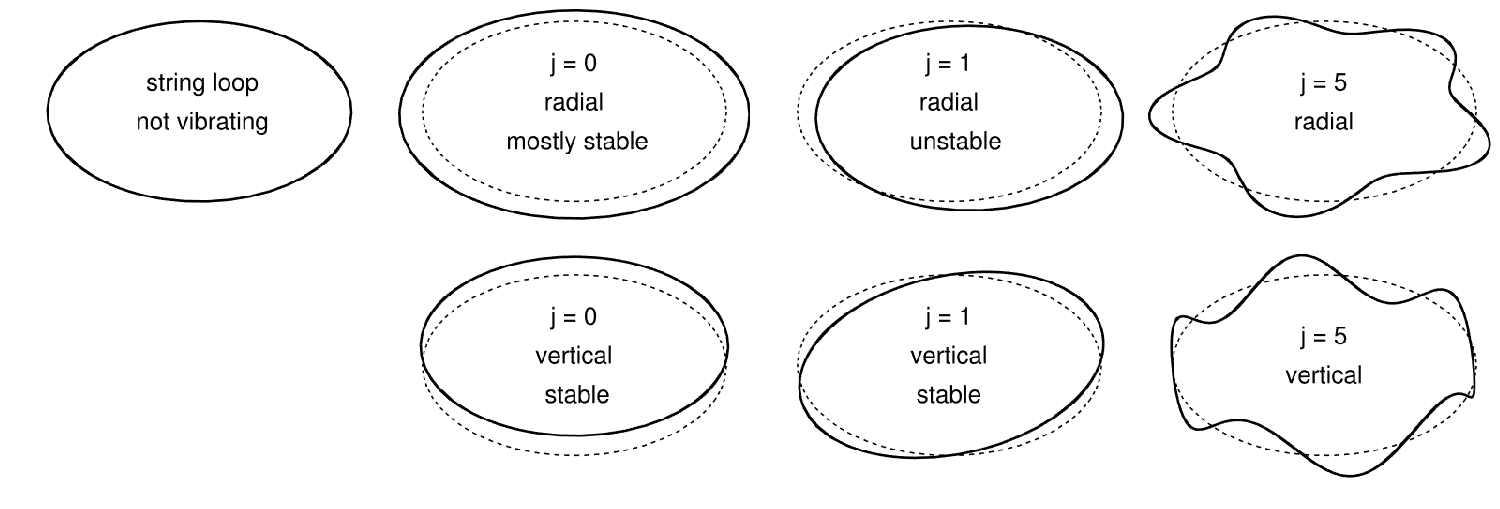}
\caption{Graphical representation of radial and vertical modes for perturbed circular string loop. Famous ringworld instability can be seen as 1st radial mode ($j=1$).
\label{figModes}
}
\end{figure*}

Based on work \cite{Nat-Que-Leo:2018:CQG:} we examine string loop vibrations around an equilibrium configuration and its stability against linear perturbations around a Schwarzschild black hole. Due to its relative simplicity, we have chosen the Schwarzschild black hole as just a starting point to a more general Kerr case. A lot of observed black holes are not rotating rapidly and Schwarzschild solution can be used as a good approximation. The string loop can orbit around the black hole without influencing the black hole rotation. String loop is a testing object only, without any direct link to make the black hole rotate. 

As established in \cite{Nat-Que-Leo:2018:CQG:}, the string loop around a Schwarzschild black hole is always stable under the polar perturbations, while it is altogether unstable under the equatorial perturbations, due to the instability of the first radial mode. We concentrate on some points that were not discussed in \cite{Nat-Que-Leo:2018:CQG:}, namely on studying equatorial perturbations for different modes separately, and we show that some higher vibrational modes are still stable for a special choice of string parameters.

\subsection{Elastic law and speeds of sound}

The string worldsheet is described by the spacetime coordinates $X^{\alpha}(\tau,\lambda)$ with $\alpha = 0,1,2,3$, given as functions of two worldsheet coordinates: $\tau$ parameterizes string evolution, $\lambda$ is running along the string. We assume that the parameter $\lambda$ is the arclength in the string unstretched configuration. The embedding $X$ induces a metric
\beq
      h_{ab}= g_{\alpha\beta}X^\alpha_{\der a}X^\beta_{\der b}
\eeq
on the worldsheet, where $g_{\alpha\beta}$ is the background metric, $\Box_{\der a} = \partial \Box /\partial a$ and $a,b\in\left\{\tau,\lambda\right\}$. The dynamic of the string is described by the action
\beq
 S = \int \mathcal{L} \, \dif \tau \dif \lambda. \label{katolicka_akce}
\eeq

Following \cite{Nat-Que-Leo:2018:CQG:}, we characterize the string by two parameters: energy density $\rho$ and pressure $p$ (since the worldsheet is two-dimensional, this pressure is actually a force -- tension or compression of the string).

For an elastic string whose internal energy density $\rho$ depends only on its stretching:
\begin{equation} \label{rho}
\rho = F(\sigma^2)\,\,,
\end{equation}
where $|\sigma|$ is the factor by which the string is stretched according to an observer comoving with it, the Lagrangian density is (see for instance \cite{Ben:1985:IJTP:, Kij-Mag:1992:JGP:})
\begin{equation} \label{Lagrangian}
\cL = F(\sigma^2) \sqrt{-h},
\end{equation}
where $h \equiv \det(h_{ab})$. Note that the stretching is related to the induced metric by
\begin{equation}
\sigma^2 = \frac{h}{h_{\tau\tau}}\,\,.
\end{equation}

The choice of the so-called {\em elastic law} $\rho = F(\sigma^2)$, corresponding to a particular kind of elastic string, defines also string pressure $p$ as
\begin{equation} \label{p}
p = -2 \sigma^2 F'(\sigma^2) - F(\sigma^2) \,\, .
\end{equation}
The relation between the energy density $\rho$ and the string pressure $p$, which is called an equation of state, is therefore determined by the internal structure of the elastic string. It can also be described by the following two parameters: the physical speed of sound for longitudinal waves
\begin{equation} \label{c^2}
c = \sqrt{\frac{dp}{d\rho}}\,\,,
\end{equation}
having the same expression as the speed of sound for a perfect fluid (see for instance \cite{Chr:2007:Book,Natario:2014:elastic:}), and the speed of sound for transverse waves
\begin{equation}
s = \sqrt{-\frac{p}{\rho}}\,\,,
\end{equation}
which generalizes the well-known classical result.

The speeds of sound $c$ and $s$ should be real. Therefore,
\begin{equation}
\frac{dp}{d\rho}\geq 0 \,,\;\;p \leq 0\,,
\end{equation}
are necessary conditions for the stability of the stretched string, as noted in \cite{Car:1989:PLB:} (otherwise there would exist exponentially growing modes in the limit of small wavelengths).

It is worth mentioning cases for string composition given by different equations of state (elastic law). For example, there is a "dust" string, i.e. ring created of test particles on a circular orbit. Such "dust string" is completely without tension, particles are not interacting, and hence $c=s=0$. On the other hand, there is an extremely stiff current-carrying string \cite{Jac-Sot:2009:PHYSR4:,Kol-Stu:2013:PHYSR4:} for which we have $c=1 ,\, s=0 $.
We have also a very famous Nambu-Goto string model \cite{Vil-She:1994:CSTD:} for which $c^2=-1 ,\, s=1 $.

\subsection{Equilibrium conditions}

The line element describing the spacetime of non-rotating Schwarzschild BH with mass $M$ is given as
\beq
    \d s^2 = -f(r) \d t^2 + f^{-1}(r) \d r^2 + r^2(\d \theta^2 + \sin^2\theta \d \phi^2), \label{SCHmetric}
\eeq
where the function $f(r)$ takes the form
\beq
	f(r) = 1 - \frac{2 M}{r}.
\eeq

As in \cite{Nat-Que-Leo:2018:CQG:}, we consider an axially symmetric elastic string loop rotating in equilibrium at a constant radius $r$ in the equatorial plane with angular velocity $\omega$,  described by an embedding
\begin{equation} \label{embed}
\begin{cases}
t(\tau,\lambda)=\tau \\
r(\tau,\lambda) = r \\
\theta(\tau,\lambda) = \frac{\pi}2 \\
\varphi(\tau,\lambda) = \omega \tau +  k \lambda
\end{cases},
\end{equation}
where $r$, $\omega$ and $k$ are constants and $\lambda \in \left[0,\frac{2\pi}{k}\right]$. Here
\begin{equation}
\frac1k = r_0  \,\, ,
\end{equation}
where $r_0$ is the radius of the unstretched string loop in flat spacetime, according to the definition of $\lambda$.
Note that the rotating string loop radius relates to the unstretched radius by the combined effects of stretching and length contraction as
\begin{equation} \label{R/R_0}
\frac{r}{r_0} = |\sigma| \sqrt{1-v^2}\,\,,
\end{equation}
where the velocity $v$ of the string with respect to the static observer is given by
\begin{equation} \label{velocity}
v = \frac{r \omega}{\sqrt{1-\frac{2 M}{r}}}\,\,.
\end{equation}

The equations of motion for the rotating string loop in the considered case lead to equation \cite{Nat-Que-Leo:2018:CQG:}
\begin{equation} \label{Kerr_eq}
\frac{p (M-r)}{r-2 M}=\frac{(p+\rho ) \left(M-r^3 \omega^2\right)}{r^3\omega^2-r+2 M}\,\,,
\end{equation}
relating the four unknown quantities $v$, $\omega$, $r$ and $\sigma$, with $r_0$ being a known and fixed parameter of the loop.


\begin{figure*}
\includegraphics[width=\hsize]{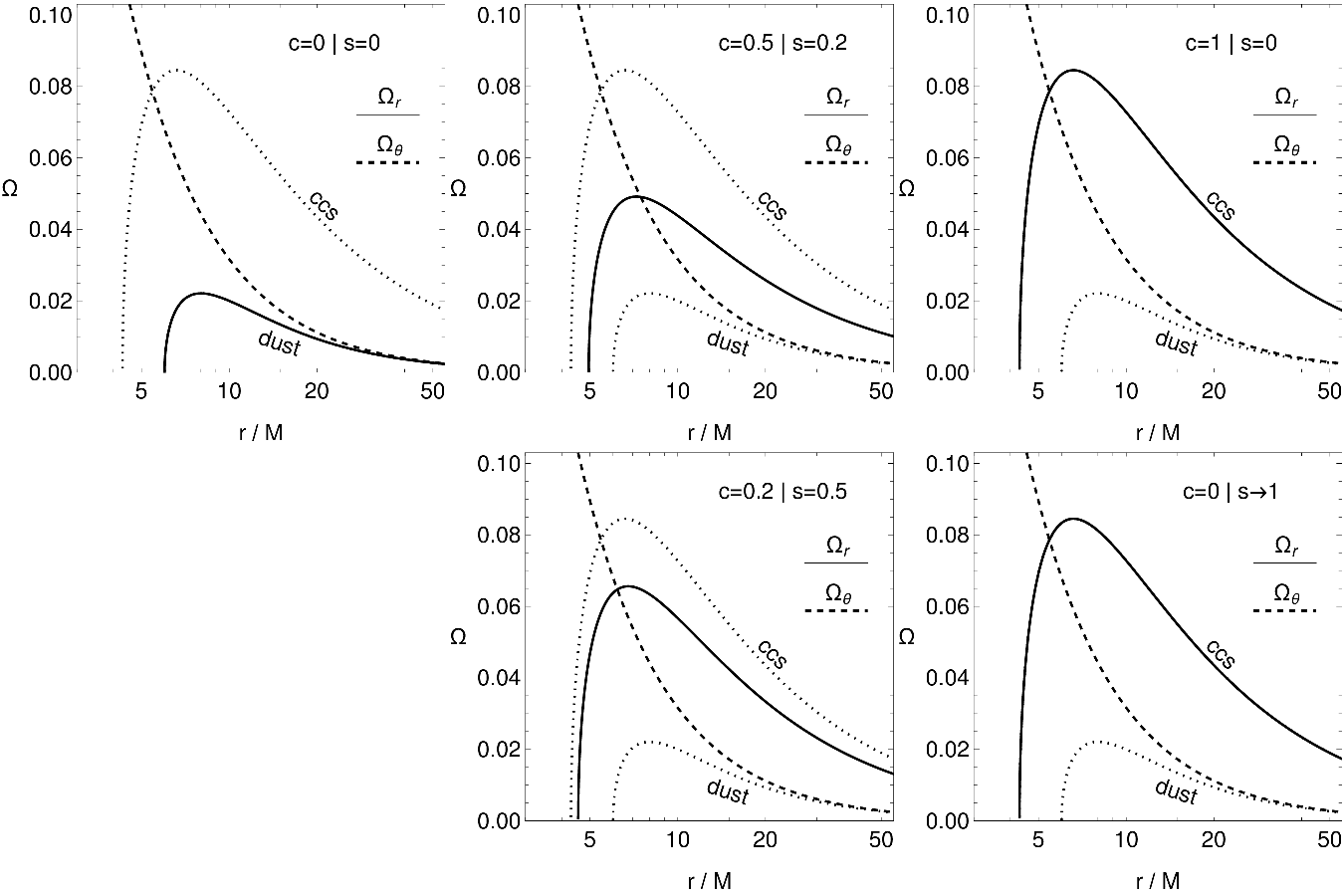}
\caption{ Radial profiles of the sting loop fundamental frequencies measured by distant measured  $\Omega_\mir$ and $\Omega_\mit$ for different string composition. We compare frequencies with two limiting cases: the case of particle motion $c=s=0$ ("dust" loop) with a current carrying string ("ccs" loop) $c=1,s=0$ (same as $c=0, s\rightarrow1$ case).
\label{figFreq}
}
\end{figure*}

\section{String loop vibrations and their stability}

We now assume that the string loop has well-defined longitudinal and transverse speeds of sound $c$ and $s$ (which is a necessary condition for stability, as noted in \cite{Car:1989:PLB:}), herewith $c,s \leq 1$, and also satisfies the weak energy condition
\begin{equation}\label{weak}
\rho \geq 0 \text{ and } \rho + p \geq 0.
\end{equation}
In particular, the weak energy condition (\ref{weak}) is satisfied by such string models as "warm" cosmic string model, non-prestressed strings with constant longitudinal speed of sound (ranging from the so-called rigid string to the incoherent, or dust, string) and strings with constant transverse speed of sound (including the Nambu-Goto string). Condition (\ref{weak}) can be violated when some exotic matter is included but this could lead to nonphysical sound velocities.
{
We first introduce, following \cite{Nat-Que-Leo:2018:CQG:}, the decomposition of general vibrations into vibrational modes, and later we treat the vibrational frequencies and stability of vibrational modes.
}

\subsection{Vibrational modes}

Analysis of the linear stability of the rotating string loops around an equilibrium configuration (satisfying Eq. \eqref{Kerr_eq})
\begin{equation} \label{embed2}
	\begin{cases}
		t(\tau,\lambda)=\tau \\
		r(\tau,\lambda) = r + \delta r(\tau, \lambda)\\
		\theta(\tau, \lambda)=\frac{\pi}{2}+\delta \theta(\tau, \lambda) \\
		\varphi(\tau,\lambda) = \omega \tau +  k \lambda + \delta \varphi(\tau,\lambda)
	\end{cases}
\end{equation}
leads to the linearized equations of motion \cite{Nat-Que-Leo:2018:CQG:}
\begin{equation} \label{motion_linear1}
A \, \delta r'+B \, \delta \varphi ''+ C \, \delta \dot{r}+ D \, \delta \dot{\varphi}' + E \, \delta \ddot{\varphi}=0 \,\, ,
\end{equation}
\begin{equation} \label{motion_linear2}
F \, \delta r + G \, \delta \varphi'+H \, \delta r''+I \, \delta \dot{\varphi}+ J \, \delta \dot{r}'+L \, \delta \ddot{r}=0 \,\, ,
\end{equation}
\begin{equation} \label{motion_linear3}
N\, \delta \theta + H\, \delta \theta''+ J\, \delta \dot{\theta}'+ L\, \delta \ddot{\theta} =0 \,\, ,
\end{equation}
with the coefficients depending on $r$, $M$, $c$, $s$ and given explicitly in Appendix~\ref{appendix1}.  Since the string loop is closed, the perturbations can be Fourier expanded.
Looking for solutions proportional to $e^{i \Omega \tau}$, one obtains that the frequencies $\Omega$ of the $j$-th mode are the roots of the corresponding characteristic polynomial: for the equatorial perturbations, they can be found from the equation \cite{Nat-Que-Leo:2018:CQG:}
\beq \label{polynomialequ-eq}
\tilde{p}_j(\Omega)=0,
\eeq
where
\begin{widetext}
\bea \label{polynomialequ}
 \tilde{p}_j(\Omega)&=\frac{1}{E L}\left[ k^2 j^2 ( A G -B F + B H k^2 j^2) +
k j \left( C G + A I -D F + (D H + B J) k^2 j^2 \right) \Omega  \right. \nonumber\\ &\left. +\left(  C I -E F +
k^2 (E H + D J + B L) j^2 \right) \Omega^2  +
k (E J + D L) j \, \Omega^3 + E L\, \Omega^4\right] ,
\eea
\end{widetext}
while for the polar perturbations, they can be found from the equation \cite{Nat-Que-Leo:2018:CQG:}
\beq
\tilde{q}_j(\Omega)=0,
\eeq
where
\beq \label{polynomialpol}
\tilde{q}_j(\Omega)=L\, \Omega^2+j k J\, \Omega -\left( N -j^2 k^2 H \right).
\eeq

\begin{figure*}
\includegraphics[width=\hsize]{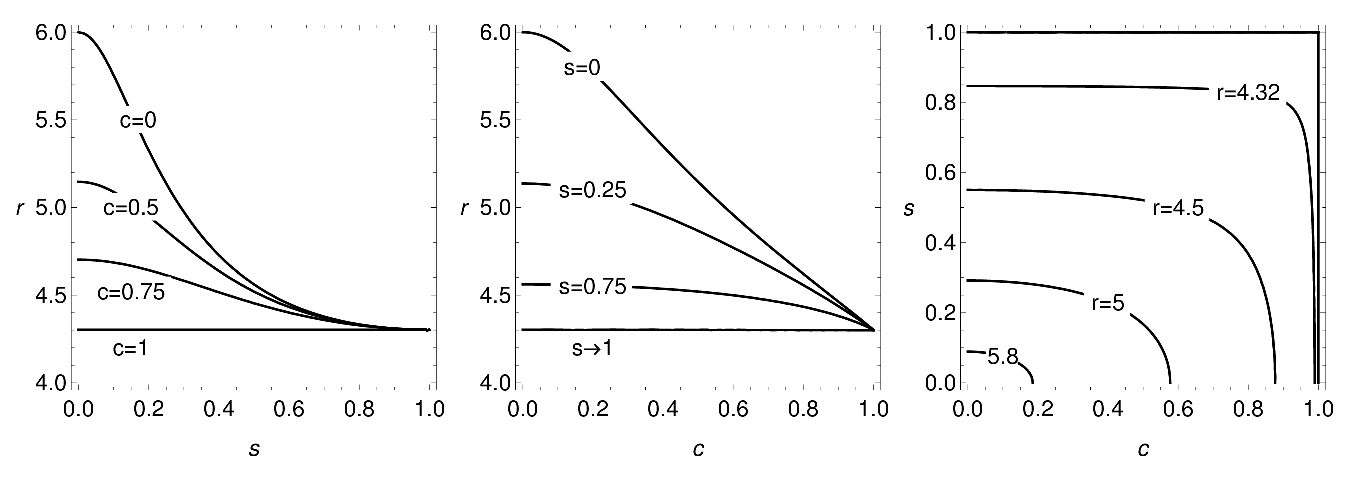}
\caption{Position of last stable circular loop radially perturbed as a function of longitudinal and traverse speed. We see the innermost stable circular orbit (ISCO) for dust string located at $r_{\rm ISCO}=6$, while current-carrying sting loop we have the innermost stable circular loop (ISCL) at $r_{\rm ISCL} \doteq 4.3$.
\label{figISCO}
}
\end{figure*}

The embedding \eqref{embed} is linearly stable under equatorial (polar) perturbations only if, for each mode $j \in \mathbb{Z}$, all the roots of $\tilde{p}_j$ ($\tilde{q}_j$) have a non-negative imaginary part. Since complex roots come in conjugate pairs, this necessary condition amounts to requiring all the roots to be real.

For the fundamental ($j=0$) mode characteristic equation for the equatorial perturbations takes the form
\bea \label{polynomialequZeroMode}
 \tilde{p}_0(\Omega)&=\left(C I -E F\right) \Omega^2  + E L \, \Omega^4=0
\eea
and for the polar perturbations take the form
\beq \label{polynomialpolZeroMode}
\tilde{q}_0(\Omega)=L\, \Omega^2- N.
\eeq

Let us first look at the polar Eq. (\ref{polynomialpolZeroMode}), which is a quadratic reduced one. It has two distinct real roots
\beq
\Omega=\pm\sqrt{\frac{N}{L}} \, ,
\eeq
since the coefficients $L$ and $N$ are positive for $r>3M$, that is outside the photon sphere, which is a natural requirement for the string loop to fulfill.

The equatorial equation (\ref{polynomialequZeroMode}) is a bi-quadratic reduced one. It has two coinciding roots $\Omega=0$ and two roots of the remaining quadratic equation
\beq
\Omega^2=\frac{E F-C I}{ E L}\, .
\eeq
Due to definition of the coefficients, whether these latter two roots are real or complex depends on the relation between longitudinal $c$ and transverse $s$ speeds of sound of the string loop under consideration.

\subsection{Frequencies of fundamental modes}

{Using the vibration theory, we can find that} the frequencies of the equatorial and polar perturbations for the fundamental mode are given correspondingly by
\begin{widetext}
\bea
\Omega^2_{\mir}=\Omega^2_{\mip} &=& \frac{1}{r^4 \left(1-s^4\right)
   (r-2 M)}\left(c^2 \left(-3 M^3+7 M^2 r-5 M r^2+r^3\right)-3 M (r-2 M)^2-M s^4 \left(12 M^2-8 M r+r^2\right)\right.\nonumber\\
    & &\left.-s^2 (r-3 M)^3-\frac{(r-3 M) \left(s^2 (r-2 M)+M\right) \left(c^2 (r-M)+s^2 (r-3 M)-2 (r-2 M)\right)^2}{M
   \left(c^2-s^2\right)-\left(1-c^2 s^2\right) (r-2 M)}\right), \label{QPOsR}\\
\Omega^2_{\mit} &=& \frac{1}{r^3} \label{QPOsT}.
\eea
\end{widetext}
In the case of "dust" string loops (with $c=0$, $s=0$) they read
\beq
 \Omega^2_{\rm \mir} = \frac{r M-6 M^2}{r^4}, \quad \Omega^2_{\rm \mit} = \frac{1}{r^3}, \label{OMd}
\eeq
while for the current carrying string loops (with $c=1$, $s=0$ or $c=0$, $s\rightarrow 1$)
\beq
\Omega^2_{\mir} = \frac{3 M^2-5 r M+r^2}{r^4},\quad \Omega^2_{\mit} = \frac{1}{r^3}. \label{OMs}
\eeq
These results coincide with the earlier results for the test particle motion (which form the "dust" string) and current-carrying string loops \cite{Kol-Stu:2013:PHYSR4:}.

There exists an innermost stable circular loop (ISCL) position, which can be calculated from the $\Omega^2_{\mir}=0$ condition. The smallest circular sting loop radius can be found t
\beq
 r_0 \rightarrow \frac{1}{2} \left(5+\sqrt{13}\right) \doteq 4.30278,
\eeq
see Fig. \ref{figISCO}. Interestingly similar value can be obtained for charged particle ISCO around magnetized BH \cite{Kol-Stu-Tur:2015:CLAQG:}.

\subsection{Higher vibration modes}

\begin{figure*}
\includegraphics[width=\hsize]{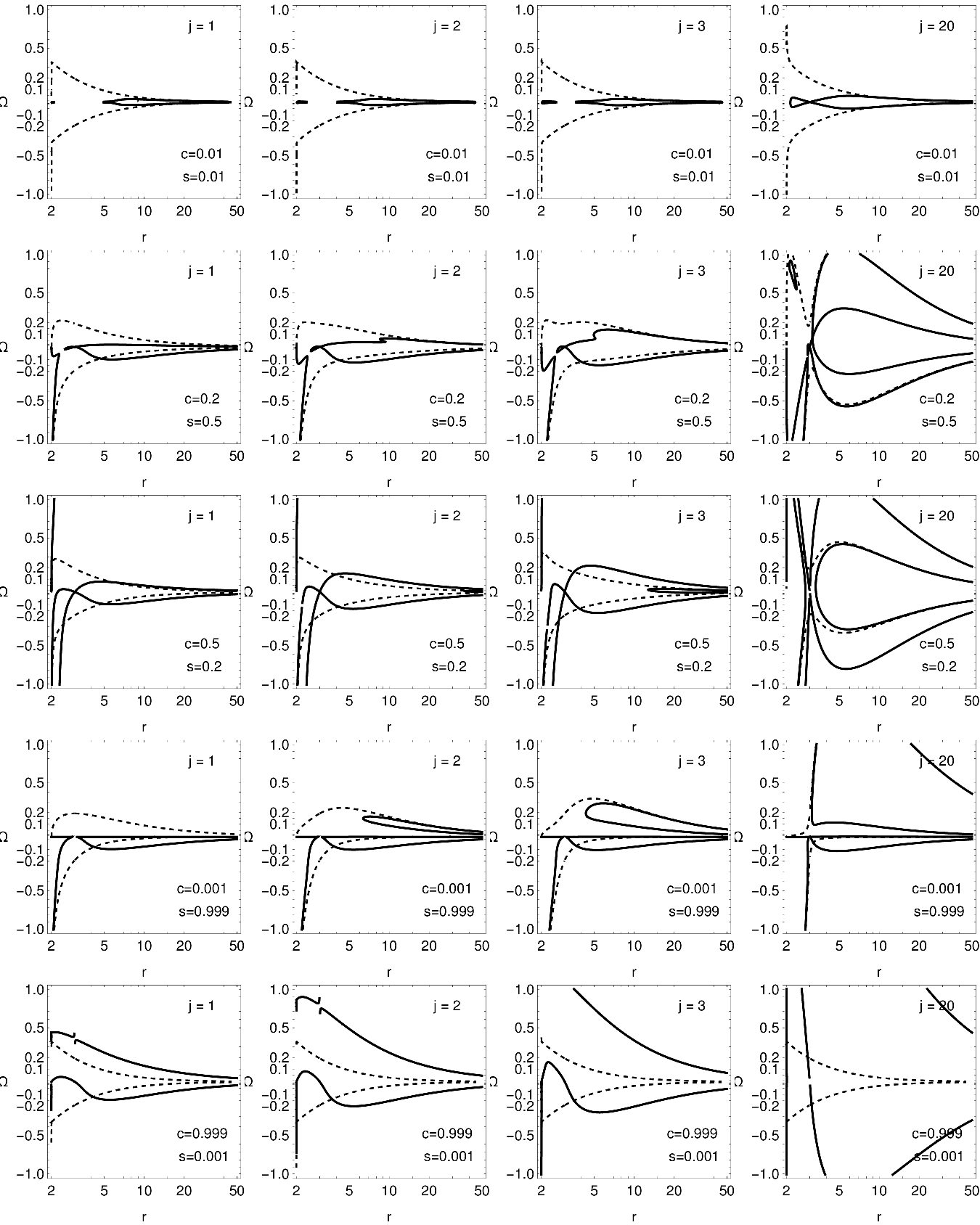}
\caption{Frequencies for higher modes of circular string loop vibrations. Thick curves are for radial perturbations $\Omega_r$, dashed for vertical one $\Omega_\theta$; here we put $M=1$. For dust string the higher modes are identical to the 0th mode (explored already in Fig.~\ref{figFreq}), for with low stiffness (almost dust) we see deviations from 0th mode only for really high modes. For stiff strings, where $c\rightarrow1$ or $s\rightarrow1$ the higher mode behavior has really complicated shape.
\label{himodes}
}
\end{figure*}

For the higher modes $j\geq1$ the characteristic equations have a more general form, containing all the powers of the variable $\Omega$. Equation (\ref{polynomialequ-eq}) for the equatorial perturbations represents a characteristic polynomial of degree 4. The number of the real roots of this equation can be inferred due to its discriminant and coefficients. In particular, if the discriminant is positive, there are four distinct roots (all real or all complex), if the discriminant is zero, there is at least one multiple root, and if the discriminant is negative, the equation has two distinct real roots and two complex roots. This means that to have four distinct real roots we need a positive discriminant. Even if we know only one real root and the discriminant is positive, we conclude that the equation has all real distinct roots. Real roots depending on the radius $r$ within the chosen range are plotted in Fig.~\ref{himodes} for different cases of $c$ and $s$ values and the number of the mode $j$. In the cases where we evidently have three or even all four real roots, as for $c=0.2$, $s=0.5$, $j=20$, $r=20$ or $c=0.001$, $s=0.999$, $j=3$, $r=10$, the discriminant is positive, as it should be, and the string loop is stable. On the other hand, for $c=0.01$, $s=0.01$, $j=2$, $r=10$ or $c=0.999$, $s=0.001$, $j=1$, $r=20$, the discriminant is negative, which means that the real roots we see on the plot are the only real roots in this case and the other two roots are complex, which means instability.

The equation for the polar perturbations with characteristic polynomial (\ref{polynomialpol}) is a quadratic one and has two distinct real roots when its discriminant is positive. Since this discriminant is positive for all the cases depicted in Fig.~\ref{himodes}, the absence of the corresponding real roots on the plots is due to the range chosen for the plots.

It is worth mentioning that in \cite{Nat-Que-Leo:2018:CQG:} the altogether stability of the circular string loop is studied, that is for all the modes at once. In particular, it has been established that the discriminant of the polar characteristic equation is always positive, and therefore the string loop is stable under the polar perturbations for all the cases. Therefore we concentrate our studies on the equatorial perturbations and investigate them for different modes $j$ separately.

In Fig.~\ref{himodes} we plot the real roots of Eq. (\ref{polynomialequ-eq}) for radial perturbations depending on the radius $r$ (thick curves) with different mode numbers $j$ for some values of the string speeds of sound in between the limiting cases of the dust string ($c=s=0 $) and current-carrying string ($c=1, s=0 $). As can be seen from the second and the third rows of the plots, for the strings with the speeds of sound $c=0.2,\,s=0.5$ and $c=0.5,\,s=0.2$ for the mode number $j=2$ or $j=3$ we already can distinguish three distinct real roots for some values of the radius $r$. This means stability for these values of the radius, which is also confirmed by the positivity of the discriminant in these cases. For the mode number $j=20$ we can see all four distinct real roots, and the range of the stability in $r$ is increasing rapidly as the mode number is growing.
The case of the string with the speeds of sound $c=0.001,\,s=0.999$ presented in the fourth row of the plots is similar: for the mode number $j=2$ we already can see three distinct real roots for some interval of the radius $r$ values, and the range of the stability in $r$ is increasing with the mode number growing.

In the near-limiting cases $c=0.01,\,s=0.01$ and $c=0.999,\,s=0.001$ (the first and the last rows of the plots) we can see at most two distinct real roots and since the discriminant turns out to be negative for these cases, this means instability up to the mode number $j=20$ and even further, with stability emerging after $j=60$ for the almost dust string ($c=0.01,\,s=0.01$) and after $j=300$ for the almost current-carrying string ($c=0.999,\,s=0.001$).

Note that for the first mode, we observe instability in all considered cases of string speeds of sound, since we can see at most two distinct real roots, which corresponds to the negativity of the discriminant in these cases.

\begin{figure}
\includegraphics[width=0.77\hsize]{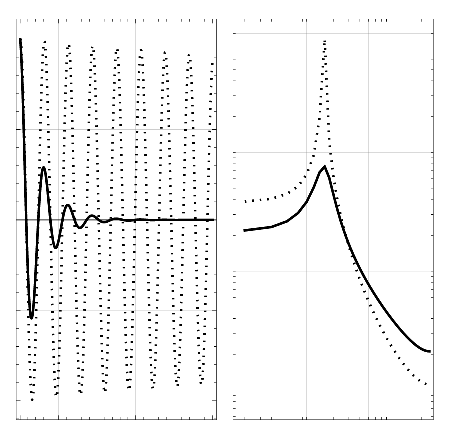}
\caption{Signal generated by a harmonic oscillator with complex frequency for Q-factor $Q=250$ (dotted) and $Q=2.5$ (solid) from the point of time (left) and frequency (right, Log-Log plot) domain.
\label{figQfactor}
}
\end{figure}

\subsection{Vibrations quality factor}

The frequencies $\Omega$ as the roots of eq. (\ref{polynomialequ-eq}) can be generally complex. Real parts $\rm{Re}(\Omega)$ are responsible for the mode harmonic oscillations, while imaginary parts $\rm{Im}(\Omega)$ are making the mode unstable - the mode will exponentially grow for $\rm{Im}(\Omega)<0$, or decay for $\rm{Im}(\Omega)>0$.
{
We can define the parameter giving the quality of oscillations related to its instability using the standard Q-factor \cite{Ing-Mot:2019:NAR:}
\beq
 Q \equiv \frac{\rm{Re}(\Omega)}{2 \, \rm{Im}(\Omega)} = \frac{\nu_0}{2\delta}, \label{qfactor}
\eeq
which will describe the relationship between the initial energy stored within the resonator and the energy dissipated during a single radian of the oscillation cycle. Such a definition is similar to the ratio of a resonator's central frequency $\nu_0$ to its bandwidth $\delta$,
which describes how narrowly peaked the signal is, see Fig.~\ref{figQfactor}.
}

\begin{figure*}
\includegraphics[width=\hsize]{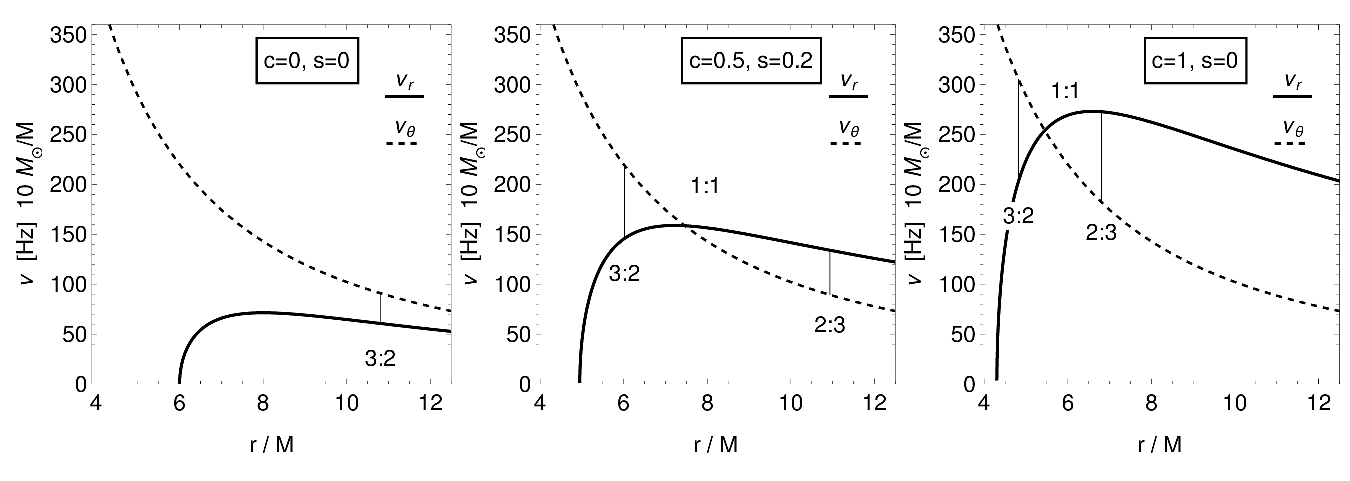}
\caption{\label{rezQPO}
Frequencies $\nu_\mit, \nu_\mir$ of small harmonic oscillations measured by static distant observers are given for the string loop around the Schwarzschild black hole. The positions of the resonant radii are also presented.}
\end{figure*}

The string loop vibration problem is reduced to the eigenvalue problem and as such is similar to the problem of finding the quasi-normal modes of the compact objects. In both cases, the solutions to the problem form a discrete infinity of complex eigenfrequencies, labeled by the mode number $j$. The real part of the solution represents the frequency of the oscillations and the imaginary part is responsible for the damping rate of the signal. Obviously in the case when the imaginary part causes an exponential growth of the corresponding eigenfunction, the object under consideration turns out to be unstable under perturbations with a particular mode number. The quasi-normal ringing for different black hole and wormhole spacetimes has been studied by some of us in  \cite{Chu-Stu:2020:CQG:,Chu:2019:EPJC:}. Here we concentrate on the possible role of string loop oscillatory motion for another astrophysically very important case of oscillations observed in accretion disk orbiting BHs or neutron stars \cite{Stu-Kot-Tor:2013:ASTRA:}.

\section{Fitting HF QPOs}

The presented string loop model is motivated by a simplified model for magnetic flux tube dynamic around BH \cite{Sem-Ber:1990:ASS:}. Such linear plasma structures are prone to various instabilities, as we have seen in our string loop model as well. Moreover, it is interesting that there is a connection between current carrying string loop vibration and frequency observed for particle motion in strong magnetic field \cite{Kol-Stu-Tur:2015:CLAQG:}. It is then attractive to explore string loop vibrations and apply the calculated frequencies to QPOs observed in microquasars.

Dimensional harmonic oscillation frequencies $\Omega_{\beta}$, related to the distant static observer, should be expressed in the physical units using conversion factor $c^3/GM$. Then the frequencies of the string loop radial and latitudinal harmonic oscillations $\nu_{r},\nu_{\theta}$ measured by the distant static observers are given by
\beq
 \nu_{\beta} = \frac{1}{2\pi} \frac{c^3}{GM} \, \Omega_{\beta}. \label{freq_units}
\eeq

\begin{table}[!h]
\begin{center}
\begin{tabular}{c l l l}
\hline
Source & GRO 1655-40 & XTE 1550-564 & GRS 1915+105 \\
\hline \hline
$ \nu_{\rm U}$ [Hz] & $447${\lin}$453$ & $273${\lin}$279$ & $165${\lin}$171$ \\
$ \nu_{\rm L}$ [Hz] & $295${\lin}$305$ & $179${\lin}$189$ & $108${\lin}$118$ \\
$ M/{M}_\odot $ & $6.03${\lin}$6.57$ & $8.5${\lin}$9.7$ & $9.6${\lin}$18.4$ \\
$ a $ & $0.65${\lin}$0.75$ & $0.29${\lin}$0.52$ & $0.98${\lin}$1$ \\
\hline
\end{tabular}
\caption{Observed twin HF QPO data for the three microquasars, and the restrictions on mass and spin of the black holes located in them, based on measurements independent of the HF QPO measurements given by the spectral continuum fitting \cite {Sha-etal:2006:ApJ:,Rem-McCli:2006:ARAA:}.} \label{tab1}
\end{center}
\end{table}


\begin{figure*}
\includegraphics[width=\hsize]{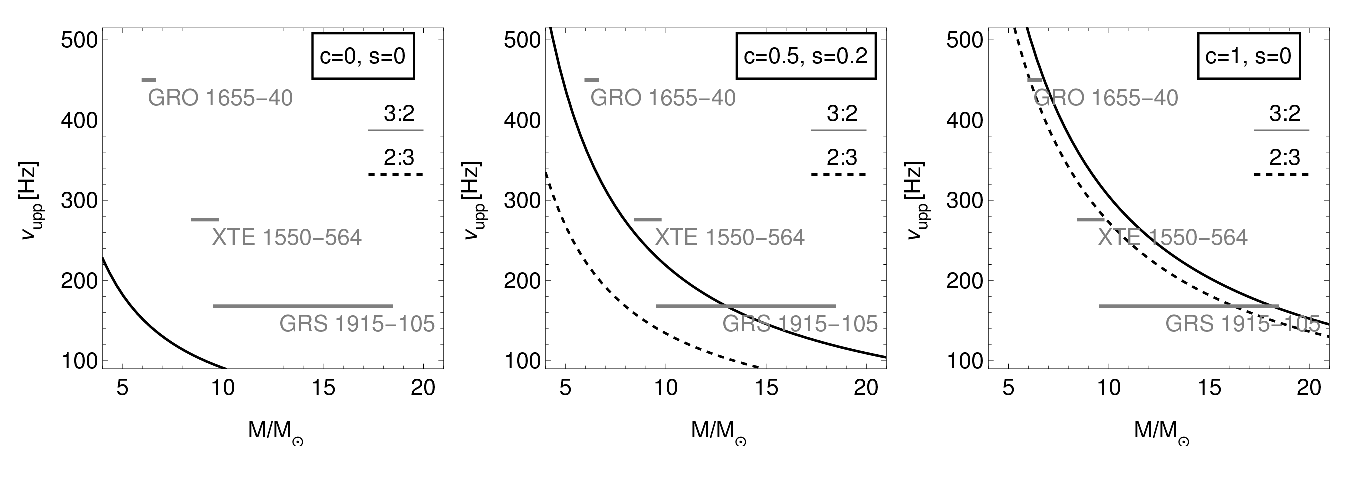}
\caption{\label{QPOfit}
String loop upper oscillation frequency $\nu_{\rm U}$ at the 3:2 or 2:3 resonance radii as a function of central BH mass (in solar mass units), see eq. (\ref{freq_units}).
 Mass-frequency limits, obtained from complementary continuum spectra observations of the three microquasars XTE~1550-564, GRO~1655-40, and GRG~1915-105, are plotted as gray thick lines. As we can see, we are able to fit observed QPOs frequencies using epicyclic resonance string loop model only for sufficiently stiff $c,s>0.3$ string equation of state
}
\end{figure*}

The quasi-harmonic character of the string loop dynamics trapped in a toroidal region around a given radius in the black hole equatorial plane suggests an interesting astrophysical application related to the HF QPOs observed in the LMXB systems containing a black hole in binary systems, or in active galactic nuclei. Some of the~HF~QPOs come in pairs of the~upper and lower frequencies ($\nu_{\mathrm{U}}$, $\nu_{\mathrm{L}}$) of twin peaks in the~Fourier power spectra. Since the peaks of high frequencies are close to the orbital frequency of the marginally stable circular orbit representing the inner edge of Keplerian discs orbiting black holes (or neutron~stars), the strong gravity effects must be relevant in explaining HF~QPOs \cite{Tor-etal:2005:ASTRA:}. Here we focus on the case of well-known microquasars, summarized in Table I.

The~twin peak HF~QPOs and their the~3\,:\,2 ratio has been pointed first out in \cite{Klu-Abr:2000:ASTROPH:} suggested that these QPOs should have rational ratios, because of the~resonances in oscillations of nearly Keplerian accretion disks; see also \cite{Ali-Gal:1981:GRG:}. It seems that the~resonance hypothesis is now well supported by observations and that the~3\,:\,2 ratio ($2\nu_{\mathrm{U}} = 3\nu_{\mathrm{L}}$) is seen most often in the twin HF QPOs in the~LMXB containing black holes (microquasars).

We will assume applicability of the parametric resonance, discussed in \cite{Lan-Lif:1969:Mech:}, focusing attention to the case of the frequency ratios $\nu_{\mit}:\nu_{\mir} = 3:2$ or $\nu_{\mit}:\nu_{\mir} = 2:3$, as the observed values of the twin HF QPO frequencies for GRO 1655-40, XTE 1550-564 and GRS 1915+105 sources show clear ratio
\beq
 \nu_{\rm U} : \nu_{\rm L} = 3 : 2
\eeq
for the upper $\nu_{\rm U}$ and lower $\nu_{\rm L}$ frequencies, see Tab. \ref{tab1}.

We use the epicyclic resonance QPOs model that is modified to string loop vibrations and identify directly the frequencies $\nu_{\rm U}, \nu_{\rm L}$ with $\nu_{\mit}, \nu_{\mir}$ or $\nu_{\mir}, \nu_{\mit}$ frequencies of the oscillating string. In contrast to the standard resonance epicyclic model based on uncharged particle dynamic, the oscillating string loop around \Schw{} black hole allows both frequency ratios
\beq
  \nu_{\mit} : \nu_{\mir} = 3 : 2, \quad \nu_{\mit} : \nu_{\mir} = 2 : 3.
\eeq
Since $ r_{3:2} < r_{2:3} $, see Figs. \ref{rezQPO}, we call the first resonance radius, where ${\nu_\mit:\nu_{\mir}=3:2}$, the inner one, and the second resonance radius, where ${\nu_\mit:\nu_{\mir}=2:3}$, the outer one. Note that for the oscillating string loop the ${1:1}, {1:2}, {2:1}$, or other, resonant frequency ratios can enter the play. There could be also resonance in just one radial or vertical direction only, but between different individual modes of the vibrating string.

The procedure of fitting the string loop oscillation frequencies to the observed frequencies is presented in Fig. \ref{QPOfit}, for all the three microquasars GRS~1915+105, XTE~1550-564, and GRO~1655-40. From the restrictions on the spacetime mass parameter $M$ for each of the sources, see Tab.~\ref{tab1}., we obtain simultaneous restrictions on string loop parameters $c$, $s$. As we can see from Fig.~\ref{QPOfit}, the string loop should be sufficiently stiff $c,s>0.5$ to fit observed HF~QPOs using the epicyclic resonance model. So far we have not considered the source rotation that can modify the radial profiles of the vertical and horizontal frequencies -- the effect of the black hole rotation is going to be studied in future work, but our preliminary results show that the fitting can be done even for the extremely fast rotating microquasar GRS~1915+105.

There are also observational restrictions on QPOs Q-factor, i.e. the harmonic oscillation should last long enough to have a strong peak in frequency spectra. As pointed out in \cite{Ing-Mot:2019:NAR:}, the Q-factor for BH QPOs should be typically between 5 and 10 or higher. The instability of the string loop 1st radial mode (ring-world instability) restricts harmonic oscillations and Q-factor dramatically. As we already discussed in the previous section, string loop higher modes have generally complex frequency and the related Q-factor is low ($Q\sim2.5$). Although there are some stable higher modes, the 1st radial mode is always unstable, and only for $c,s<0.01$ we do have the Q-factor strong enough to fit observed QPOs. Parameter $c,s\sim0.01$ can give us restrictions on maximal plasma (magnetic flux tube) stiffness within our string loop model.

\begin{figure*}
\flushleft\includegraphics[width=0.67\hsize]{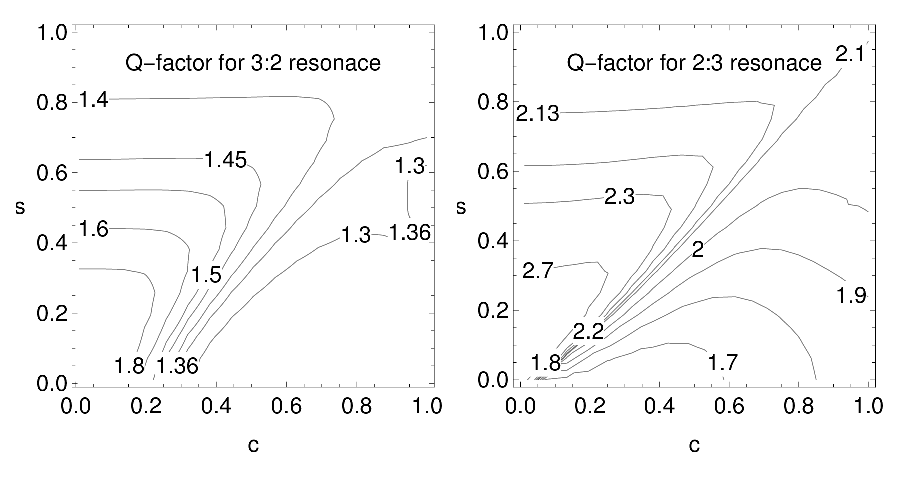}
\caption{\label{qfactor_rez}
Stability of 3:2 and 2:3 resonance for string loop zeroth mode. The Q-factor is calculated from ratio of upper resonant frequency given by stable (real) zeroth mode to the unstable (imaginary) value of first radial mode, see definition (\ref{qfactor}). The most stable oscillations (highest Q-factor) can be found for small $c,s$.
}
\end{figure*}

Unfortunately for $c,s\sim0.01$ the string loop vibrational frequencies are unable to fit observed HF~QPOs using the presented simple epicyclic resonance model. More complex sting loop QPOs model with different frequency combination should be used or one should consider resonances between basic and higher modes in separate radial and vertical directions.

\section{Conclusions} \label{kecy}

Based on work \cite{Nat-Que-Leo:2018:CQG:} we examined string loop vibrations around an equilibrium configuration and its stability for different radii around a Schwarzschild black hole in linear approach. While in \cite{Nat-Que-Leo:2018:CQG:} the altogether stability of the circular string loop was already studied, that is for all the modes at once, here we provided frequency analysis for different modes of vibrations, and also for different string loop composition, in dependence on speeds of sound $c,s$ within the string. As established in \cite{Nat-Que-Leo:2018:CQG:}, the string loop around a Schwarzschild black hole is always stable under the polar perturbations, while it is altogether unstable under the equatorial perturbations, since the 1st radial mode is always unstable (ringworld instability). We study equatorial perturbations for different modes separately and show that some higher vibrational modes are still stable but only for a special choice of $c,s$.

{

As we already pointed out, there is a connection between one-dimensional plasma structures (magnetic flux tubes) and relativistic string which is based on the similarity of their equation of motion \cite{Spruit:1982:Sol:,Sem-Ber:1990:ASS:}. Contrary to the dust particle orbiting the BH freely along a circular geodesic, the string loop circular structure (plasma flux tube) has some internal tension. On the other hand, the thick accretion torus formed around the BH needs some pressure force to maintain its size and if the torus pressure is removed, the torus will deflate into a circular orbit as well. Hence one can see the presented string loop model as another accretion disk model, approached from different side and introducing stiffness (tension) into play. Both thick torus and string loop are prone to radial instabilities which have geometrical reasons in both cases: for the thick torus it is Papaloizou–Pringle instability \cite{Pap-Pri:1984:MNRAS:} and for string loop it is ring-world instability \cite{Nat-Que-Leo:2018:CQG:}.

Our article is basically a theoretical study of a structure with tension (string loop) orbiting in the accretion disk around a nonrotating BH. However, the string loop vibration frequencies and stability of different vibrational modes (Q-factor) can be easily confronted with observed QPO signals. A single test particle does not have the strength to produce observed X-ray flux change and some extended object (accretion torus, string loop) or some plasma collective phenomena are needed. Since the observed QPOs have a limited lifespan, we need a mechanism to create the vibration and to let them die - and for this effect the sting loop radial instability is convenient. In this work, we demonstrated, that zero mode frequency of string loop oscillations, and resonances between vertical and radial mode can be used for fitting observed HF~QPOs in X-ray microquasars quite well using the epicyclic resonance model. However, the string loops are unstable throughout ringworld instability in the 1st radial mode, with quite a short lifetime for high sound speeds $c$ and $s$. Larger and more reasonable Q~factors (more stable oscillations) can be obtained for strings with low stiffness $c,s<0.1$ (dust-like). Observed black hole HF~QPOs have Q~factor typically between 5 and 10 or higher \cite{Ing-Mot:2019:NAR:} and such Q~factor condition restricts string models to weak dust string only. Unfortunately for such almost dust string the HF~QPOs frequency fit using the epicyclic resonance model is not working in the pure Schwarzschild spacetime. For almost dust string, the frequencies are very close to those obtained for test particles, and a more complex oscillation model should be used. We can summarize the results of our string loop model in the Schwarzschild geometry for different string equations of state: when we are able to fit the frequency data, the Q-factor is too low; and when a good Q-factor is obtained, frequencies will not match the QPOs data. Therefore, modifications of the string equations of state are needed, or modification of the background spacetime (Kerr geometry) could be useful.

It is worth noting that both pressure and tension effects can be combined into a single model of thick accretion torus with a toroidal magnetic field as introduced by Komissarov \cite{Kom:2006:MNRAS:} and applied e.g. in \cite{Pug-Mon:2018:MNRAS:}. One of the results of this study is that when tension is included in accretion disk structure the vibrational frequencies will be increased, see how string loop frequencies are increased with string tension and compare them with free test particle frequencies. This is the opposite effect of thick torus pressure forces which will make frequencies go down as the structure becomes thicker \cite{Str-Sra:2009:CQG:}. One can conclude that the frequencies of structure vibrations are going up as the object becomes stiffer, and going down when the object becomes softer, which is in accordance with our daily experience. It will be interesting to see if the stiffer accretion torus with a toroidal magnetic field will be more resilient to various instabilities, which are very common in GRMHD simulations of oscillating structures. In some of our future work, we would like to give realistic estimates on string loop speeds of sound $c,s$ using plasma flux tube model \cite{Spruit:1982:Sol:}. Also, the closed string loop evolution for any general shape beyond the linear perturbation model using numerical integration of the equation of motion could be fruitful with application to magnetic flux tube dynamic \cite{Sem-Ber:1990:ASS:}.

}


\section*{Acknowledgments}

The work is supported by the Research Centre for Theoretical Physics and Astrophysics, Institute of Physics, Silesian University in Opava, and the GA{\v C}R~\mbox{23-07043S} project.

\appendix

\section{Characteristic polynomials coefficients} \label{appendix1}

The coefficients of Eqs. \eqref{motion_linear1}--\eqref{motion_linear3} are \cite{Nat-Que-Leo:2018:CQG:}:

\begin{widetext}
\begin{align*}
& A=- \frac{(r-M) (s^2+c^2 ) \sqrt{(r-2M)s^2+M}}{k (r-2M)^{3/2} \sqrt{r^3 \left[ (r-2M)+M s^2 \right]}}; \\
& B=- \frac{c^2 r (1-s^2) \sqrt{(r-2M)s^2+M}}{k^2 \sqrt{r-2 M} \sqrt{r^3 \left[(r-2M) +M s^2\right]}}; \\
& C=- \frac{\left[(r-2M)s^2+M \right] \left[(r-M)c^2 +(r-3M)s^2-2(r-2M) \right]}{(r-3 M) (r-2 M)^2 (1-s^2 )}; \\
& D=- \frac{2 c^2 r \left[(r-2M)s^2+M \right]}{k (r-2 M) (r-3 M)}; \\
& E=- \frac{r \sqrt{r^3 \left[(r-2M)+M s^2 \right]} \sqrt{(r-2M)s^2+M}\, \left[M (c^2-s^2)-(r-2 M) (1-c^2 s^2)\right]}{(r-2 M)^{3/2} (r-3 M)^2 (1-s^2 )}; \\
& F= \frac{ c^2\left(-3 M^3+7 M^2 r-5 M r^2+ r^3\right)-3 M (r-2M)^2-M s^4 \left(12 M^2-8 M r+r^2\right)-s^2  (r-3 M)^3}{r^3 (r-3 M) (r-2 M) (1-s^2 )}; \\
& G= \frac{(r-M) (s^2+c^2 )}{k r^2}; \\
& H=- \frac{s^2 (r-3 M) (1-s^2)}{k^2 r^2 \left[(r-2M)+M s^2\right]}; \\
& I= \frac{\sqrt{r^3 \left[(r-2M)+M s^2\right]}\, \sqrt{(r-2M)s^2+M} \left[c^2(r-M)+s^2 (r-3M)-2(r-2M) \right]}{r^2 \sqrt{r-2 M}\, (r-3 M) (1-s^2)}; \\
& J= - \frac{2 s^2 r \sqrt{(r-2M)s^2+M}}{k \sqrt{r-2 M}\, \sqrt{r^3 \left[(r-2M)+M s^2\right]}}; \\
& L= \frac{r (1+s^2)}{r-3 M}; \\
& N= \frac{M (1+s^2)}{r^2 (r-3 M)} .
\end{align*}
\end{widetext}
The quantities $s$ and $c$ are taken at the particular equilibrium under consideration.



\def\prc{Phys. Rev. C}
\def\pre{Phys. Rev. E}
\def\prd{Phys. Rev. D}
\def\jcap{Journal of Cosmology and Astroparticle Physics}
\def\apss{Astrophysics and Space Science}
\def\mnras{Monthly Notices of the Royal Astronomical Society}
\def\apj{The Astrophysical Journal}
\def\aap{Astronomy and Astrophysics}
\def\actaa{Acta Astronomica}
\def\pasj{Publications of the Astronomical Society of Japan}
\def\apjl{Astrophysical Journal Letters}
\def\pasa{Publications Astronomical Society of Australia}
\def\nat{Nature}
\def\physrep{Physics Reports}
\def\araa{Annual Review of Astronomy and Astrophysics}
\def\apjs{The Astrophysical Journal Supplement}
\def\aapr{The Astronomy and Astrophysics Review}
\def\procspie{Proceedings of the SPIE}

\bibliographystyle{plain}


\begin{thebibliography}{10}

\bibitem{Ali-Gal:1981:GRG:}
A.~N. {Aliev} and D.~V. {Galtsov}.
\newblock {Radiation from relativistic particles in nongeodesic motion in a
  strong gravitational field}.
\newblock {\em General Relativity and Gravitation}, 13:899--912, October 1981.

\bibitem{Ben:1985:IJTP:}
L.~{Bento}.
\newblock {Transverse Waves in a Relativistic Rigid Body}.
\newblock {\em International Journal of Theoretical Physics}, 24(6):653--657,
  June 1985.

\bibitem{Car:1989:PLB:}
B.~{Carter}.
\newblock {Stability and characteristic propagation speeds in superconducting
  cosmic and other string models}.
\newblock {\em Physics Letters B}, 228(4):466--470, September 1989.

\bibitem{Chr:2007:Book}
Demetrios {Christodoulou}.
\newblock {\em {The Formation of Shocks in 3-Dimensional Fluids}}.
\newblock European Mathematical Society, 2007.

\bibitem{Chu:2019:EPJC:}
M.~S. {Churilova}.
\newblock {Analytical quasinormal modes of spherically symmetric black holes in
  the eikonal regime}.
\newblock {\em European Physical Journal C}, 79(7):629, July 2019.

\bibitem{Chu-Stu:2020:CQG:}
M.~S. {Churilova} and Z.~{Stuchl{\'\i}k}.
\newblock {Ringing of the regular black-hole/wormhole transition}.
\newblock {\em Classical and Quantum Gravity}, 37(7):075014, April 2020.

\bibitem{Daly:2019:APJ:}
Ruth~A. {Daly}.
\newblock {Black Hole Spin and Accretion Disk Magnetic Field Strength Estimates
  for More Than 750 Active Galactic Nuclei and Multiple Galactic Black Holes}.
\newblock {\em \apj}, 886(1):37, 2019.

\bibitem{Eatough-etal:2013:Natur:}
R.~P. {Eatough}, H.~{Falcke}, R.~{Karuppusamy}, K.~J. {Lee}, D.~J. {Champion},
  E.~F. {Keane}, G.~{Desvignes}, D.~H.~F.~M. {Schnitzeler}, L.~G. {Spitler},
  M.~{Kramer}, B.~{Klein}, C.~{Bassa}, G.~C. {Bower}, A.~{Brunthaler},
  I.~{Cognard}, A.~T. {Deller}, P.~B. {Demorest}, P.~C.~C. {Freire},
  A.~{Kraus}, A.~G. {Lyne}, A.~{Noutsos}, B.~{Stappers}, and N.~{Wex}.
\newblock {A strong magnetic field around the supermassive black hole at the
  centre of the Galaxy}.
\newblock {\em \nat}, 501:391--394, 2013.

\bibitem{Ing-Mot:2019:NAR:}
Adam~R. {Ingram} and Sara~E. {Motta}.
\newblock {A review of quasi-periodic oscillations from black hole X-ray
  binaries: Observation and theory}.
\newblock {\em New Astronomy Reviews}, 85:101524, September 2019.

\bibitem{Jac-Sot:2009:PHYSR4:}
T.~{Jacobson} and T.~P. {Sotiriou}.
\newblock {String dynamics and ejection along the axis of a spinning black
  hole}.
\newblock {\em Physical Review D}, 79(6):065029, March 2009.

\bibitem{Kij-Mag:1992:JGP:}
Jerzy {Kijowski} and Giulio {Magli}.
\newblock {Relativistic elastomechanics as a lagrangian field theory}.
\newblock {\em Journal of Geometry and Physics}, 9(3):207--223, July 1992.

\bibitem{Klu-Abr:2000:ASTROPH:}
W.~{Kluzniak} and M.~A. {Abramowicz}.
\newblock {The physics of kHz QPOs---strong gravity's coupled anharmonic
  oscillators}.
\newblock {\em ArXiv Astrophysics e-prints}, May 2001.

\bibitem{Kol-Stu:2013:PHYSR4:}
M.~{Kolo{\v s}} and Z.~{Stuchl{\'{\i}}k}.
\newblock {Dynamics of current-carrying string loops in the Kerr
  naked-singularity and black-hole spacetimes}.
\newblock {\em \prd}, 88(6):065004, September 2013.

\bibitem{Kol-Stu-Tur:2015:CLAQG:}
M.~{Kolo{\v s}}, Z.~{Stuchl{\'{\i}}k}, and A.~{Tursunov}.
\newblock {Quasi-harmonic oscillatory motion of charged particles around a
  Schwarzschild black hole immersed in a uniform magnetic field}.
\newblock {\em Classical and Quantum Gravity}, 32(16):165009, August 2015.

\bibitem{Kom:2006:MNRAS:}
S.~S. {Komissarov}.
\newblock {Magnetized tori around Kerr black holes: analytic solutions with a
  toroidal magnetic field}.
\newblock {\em \mnras}, 368(3):993--1000, May 2006.

\bibitem{Lan-Lif:1969:Mech:}
L.~D. {Landau} and E.~M. {Lifshitz}.
\newblock {\em {Mechanics}}.
\newblock Oxford: Pergamon Press, 1969.

\bibitem{Larsen:1994:CQG:}
A.~L. {Larsen}.
\newblock {Chaotic string-capture by black hole}.
\newblock {\em Classical and Quantum Gravity}, 11(5):1201--1210, May 1994.

\bibitem{Natario:2014:elastic:}
Jose {Natario}.
\newblock {Relativistic elasticity of rigid rods and strings}.
\newblock {\em arXiv e-prints}, page arXiv:1406.0634, June 2014.

\bibitem{Nat-Que-Leo:2018:CQG:}
Jos{\'e} {Nat{\'a}rio}, Leonel {Queimada}, and Rodrigo {Vicente}.
\newblock {Rotating elastic string loops in flat and black hole spacetimes:
  stability, cosmic censorship and the Penrose process}.
\newblock {\em Classical and Quantum Gravity}, 35(7):075003, April 2018.

\bibitem{Pap-Pri:1984:MNRAS:}
J.~C.~B. {Papaloizou} and J.~E. {Pringle}.
\newblock {The dynamical stability of differentially rotating discs with
  constant specific angular momentum}.
\newblock {\em \mnras}, 208:721--750, June 1984.

\bibitem{Pug-Mon:2018:MNRAS:}
D.~{Pugliese} and G.~{Montani}.
\newblock {Influence of toroidal magnetic field in multiaccreting tori}.
\newblock {\em \mnras}, 476(4):4346--4361, June 2018.

\bibitem{Rem-McCli:2006:ARAA:}
R.~A. {Remillard} and J.~E. {McClintock}.
\newblock {X-Ray Properties of Black-Hole Binaries}.
\newblock {\em \araa}, 44:49--92, September 2006.

\bibitem{Sem-Dya-Pun:2004:Sci:}
V.~{Semenov}, S.~{Dyadechkin}, and B.~{Punsly}.
\newblock {Simulations of Jets Driven by Black Hole Rotation}.
\newblock {\em Science}, 305:978--980, August 2004.

\bibitem{Sem-Ber:1990:ASS:}
V.~S. {Semenov} and L.~V. {Bernikov}.
\newblock {Magnetic flux tubes - Nonlinear strings in relativistic
  magnetohydrodynamics}.
\newblock {\em \apss}, 184:157--166, October 1991.

\bibitem{Sha-etal:2006:ApJ:}
R.~{Shafee}, J.~E. {McClintock}, R.~{Narayan}, S.~W. {Davis}, L.-X. {Li}, and
  R.~A. {Remillard}.
\newblock {Estimating the Spin of Stellar-Mass Black Holes by Spectral Fitting
  of the X-Ray Continuum}.
\newblock {\em \apjl}, 636:L113--L116, January 2006.

\bibitem{Spruit:1982:Sol:}
H.~C. {Spruit}.
\newblock {Propagation Speeds and Acoustic Damping of Waves in Magnetic Flux
  Tubes}.
\newblock {\em Solar Physics}, 75(1-2):3--17, January 1982.

\bibitem{Str-Sra:2009:CQG:}
Odele {Straub} and Eva {{\v{S}}r{\'a}mkov{\'a}}.
\newblock {Epicyclic oscillations of non-slender fluid tori around Kerr black
  holes}.
\newblock {\em Classical and Quantum Gravity}, 26(5):055011, March 2009.

\bibitem{Stu-Kol:2014:PHYSR4:}
Z.~{Stuchl{\'{\i}}k} and M.~{Kolo{\v s}}.
\newblock {String loops oscillating in the field of Kerr black holes as a
  possible explanation of twin high-frequency quasiperiodic oscillations
  observed in microquasars}.
\newblock {\em \prd}, 89(6):065007, March 2014.

\bibitem{Stu-Kot-Tor:2013:ASTRA:}
Z.~{Stuchl{\'{\i}}k}, A.~{Kotrlov{\'a}}, and G.~{T{\"o}r{\"o}k}.
\newblock {Multi-resonance orbital model of high-frequency quasi-periodic
  oscillations: possible high-precision determination of black hole and neutron
  star spin}.
\newblock {\em \aap}, 552:A10, April 2013.

\bibitem{Stu-etal:2020:Universe:}
Zden{\v{e}}k {Stuchl{\'\i}k}, Martin {Kolo{\v{s}}}, Ji{\v{r}}{\'\i}
  {Kov{\'a}{\v{r}}}, Petr {Slan{\'y}}, and Arman {Tursunov}.
\newblock {Influence of Cosmic Repulsion and Magnetic Fields on Accretion Disks
  Rotating around Kerr Black Holes}.
\newblock {\em Universe}, 6(2):26, Jan 2020.

\bibitem{Tor-etal:2005:ASTRA:}
G.~{T{\"o}r{\"o}k}, M.~A. {Abramowicz}, W.~{Klu{\'z}niak}, and
  Z.~{Stuchl{\'{\i}}k}.
\newblock {The orbital resonance model for twin peak kHz quasi periodic
  oscillations in microquasars}.
\newblock {\em \aap}, 436:1--8, June 2005.

\bibitem{Vil-She:1994:CSTD:}
A.~{Vilenkin} and E.~P.~S. {Shellard}.
\newblock {\em {Cosmic Strings and Other Topological Defects}}.
\newblock Cambridge University Press, January 1995.

\bibitem{Wit:1985:NuclPhysB:}
E.~{Witten}.
\newblock {Superconducting strings}.
\newblock {\em Nuclear Physics B}, 249:557--592, 1985.

\end{thebibliography}

\end{document}